\documentstyle[12pt]{article}
\begin{document}
\section*{}
\hspace*{\fill} IASSNS--HEP--95/73\\
\hspace*{\fill} hep-th/9509118\\[3ex]

\begin{center}
\bf
SELF--TRAPPING OF THE DILATON
\end{center}
\vspace{-6ex}
\section*{}
\normalsize \rm
\begin{center}
{\bf Rainer Dick}\\[0.5ex] {\small \it School of Natural Sciences, Institute
for Advanced Study\\ Olden Lane, Princeton, NJ 08540, USA \\[0.5ex] {\rm
and}\\[0.5ex]
Department of Physics,
University of Munich\\
Theresienstr.\ 37, 80333 Munich,
Germany}

\end{center}

\section*{}
{\bf Abstract}: The dilaton in
three dimensions does not roll. Witten's conjecture that duality between
theories in three and four dimensions solves the cosmological constant
problem thus may also solve the dilaton problem in string theory.
\\[3ex]

\newpage
%\section{Introduction}
Recently
Witten pointed out that theories with Bogomol'nyi saturated
solitons may be related to theories in higher dimensions in a Kaluza--Klein
type framework \cite{Wi1,Wi2}.
This idea can be motivated from classical duality considerations
which generically imply a trading between solitons and particles
in dual theories. Given the soliton--particle correspondence and the infinite
tower of equidistant solitonic excitations it seems very natural to expect that
theories with Bogomol'nyi saturated solitons are related to compactified
theories
in higher dimensions.
In this framework the particular case of dualities between
supersymmetric theories in 2+1 dimensions and non--supersymmetric
theories in 3+1 dimensions deserves special attention, since the
four--dimensional theory
might inherit the vanishing of the cosmological constant from the corresponding
three--dimensional theory\footnote{Supersymmetry as a solution to the
cosmological constant problem has been discussed
in \cite{BZ}. A very useful review and critical discussion of several
attempts to solve the problem can be found in \cite{SW}.
Early references on supergravity in 2+1 dimensions are \cite{HT,DK}.}
\cite{Wi2,Wi3}.
A recent discussion of the supersymmetric abelian Higgs model in 2+1 dimensions
coupled to supergravity confirmed this picture by showing that the soliton
spectrum in this theory is not supersymmetric \cite{BBS}.

However, one might expect difficulties with this scenario due to the
logarithmic potential
in 2+1 dimensions. In this note I would like to point out that the
infrared singularity of the dilaton in three dimensions actually helps to solve
the dilaton problem in string theory, since it implies a self--trapping of the
dilaton: In 2+1 dimensions fermions and adjoint scalars
provide sources for the dilaton which differ in sign from the
dilaton sources provided by the gauge fields which arise from
the four--dimensional metric and four--dimensional gluons. Finite
energy then implies that any local dilaton source has to be compensated
by another dilaton source somewhere else, whence the dilaton vanishes
asymptotically. In particular, the dilaton may fluctuate, but it does not roll.
Hence, in the three--dimensional scenario string theory predicts
fluctuations of low--energy couplings, but not a  monotonic evolution.

In order to make this observation quantitative, I will also
take a four--dimensional point of view like in \cite{Wi3} and
discuss the
action of the $(D-1)$--dimensional
dilaton arising through a Kaluza--Klein parametrization
of Einstein--Yang--Mills theory in $D$ dimensions.
The Kaluza--Klein {\sl Ansatz} (see e.g.\ \cite{AC})
\begin{equation}\label{KK}
G_{MN}=\Phi^{-\frac{1}{D-2}}\left(\begin{array}{cc} g_{\mu\nu}+\Phi a_\mu a_\nu
& \,\Phi a_\mu \\ \Phi a_\nu & \,\Phi\end{array}\right)
\end{equation}
yields in the gravitational sector (for the zero modes of
$\partial_{D-1}$ and up to surface terms)
\begin{equation}
\sqrt{-G}\,G^{MN}R_{MN}=\sqrt{-g}\,[g^{\mu\nu}R_{\mu\nu}-
\frac{D-3}{4(D-2)}g^{\mu\nu}\partial_\mu\ln\Phi\cdot
\partial_\nu\ln\Phi-\frac{\Phi}{4}f_{\mu\nu}f^{\mu\nu}]
\end{equation}
while reduction of the Yang--Mills field $A_M^{\alpha}\to
A_\mu^\alpha,A^\alpha$
yields
\begin{equation}
-\frac{1}{4}\sqrt{-G}\,F_{MN}^{\alpha}F^{MN}_{\alpha}=-\frac{1}{4}\sqrt{-g}\,
\Phi^{\frac{1}{D-2}}[B_{\mu\nu}^\alpha B^{\mu\nu}_\alpha
+\frac{2}{\Phi}g^{\mu\nu}
D_\mu A^\alpha \cdot D_\nu A_\alpha ]
\end{equation}
\[
B_{\mu\nu}=F_{\mu\nu}+a_\mu D_\nu A-a_\nu D_\mu A
\]

We discuss reduction of the fermion terms for even $D$, since the case of odd
$D$ is simpler, and we are mainly interested in the case $D=4$. If $\gamma_\mu$
is a basis of Dirac matrices in $D-1$ dimensions, a convenient basis of Dirac
matrices in $D$ dimensions is given by
\[
\Gamma_0 = \left(\begin{array}{cc} 0 & \,1\\ 1 & \,0\end{array}\right)\]
\[
\Gamma_j =
\left(\begin{array}{cc} 0&-\gamma_0\gamma_j\\
\gamma_0\gamma_j&0\end{array}\right)\]
\[
\Gamma_{D-1} = \left(\begin{array}{cc} 0 & \gamma_0\\ -\gamma_0 &
0\end{array}\right)
\]

The reduction of
fermions yields both inequivalent representations of the Clifford algebra
in $D-1$ dimensions.
In a gauge $E^\mu{}_{D-1}=0$ for the $D$--{\sl bein} and with a Kaluza--Klein
{\sl Ansatz}
\[
\Psi=\Phi^{\frac{1}{4(D-2)}}\left(\begin{array}{c}\psi_+\\ \psi_-
\end{array}\right)
\]
the result
reads
\begin{equation}
\sqrt{-G}\,\overline{\Psi}[E^M{}_A\Gamma^A(i\partial_M
+i\Omega_M +qA_M)-M]\Psi=
\end{equation}
\[
=
\sqrt{-g}\,[\overline{\psi}_+ e^\mu{}_a\gamma^a_+
(i\partial_\mu +i\omega_{+\mu}+qV_\mu)\psi_+
+
\overline{\psi}_- e^\mu{}_a\gamma^a_-(i\partial_\mu
+i\omega_{-\mu} +qV_\mu)\psi_-]
\]
\[
-q\sqrt{-g}\,\Phi^{-\frac{1}{2}}(\overline{\psi}_+ A\psi_+ +\overline{\psi}_-
A\psi_-)+M\sqrt{-g}\,\Phi^{-\frac{1}{2(D-2)}}(\psi_+^+\psi_- +\psi_-^+\psi_+)
\]
\[
-\frac{i}{8}\sqrt{-g}\,\Phi^{\frac{1}{2}}f_{ab}
(\overline{\psi}_+ \gamma_+^{ab}\psi_+ +\overline{\psi}_-
\gamma_-^{ab}\psi_-)
\]
with $\Omega$ and $\omega_\pm$ denoting the canonical spin connections,
$\gamma^0_\pm = \pm \gamma^0$, $\gamma^j_\pm =\gamma^j$, and
\[
V_\mu =A_\mu-a_\mu A \]
\[
V_{\mu\nu}=B_{\mu\nu}-Af_{\mu\nu}
\]

In order to evaluate the effect of the infrared divergence of the
electrostatic potential in 2+1 dimensions, we consider the energy
of a static configuration with the fermions in stationary orbits,
and in gauge $A_0=a_0=0$ (complying with $E^\mu{}_{D-1}=0$, since we employ
diffeomorphisms  which are constant along $x^{D-1}$):
\begin{equation}
\frac{{\cal H}}{\sqrt{-g}}=\frac{D-3}{4(D-2)}g^{ij}\partial_i\ln\Phi\cdot
\partial_j\ln\Phi+\frac{\Phi}{4}f_{ij}f^{ij}
+\frac{1}{4}
\Phi^{\frac{1}{D-2}}(V+Af)_{ij}^\alpha (V+Af)^{ij}_\alpha
\end{equation}
\[
+\frac{1}{2}
\Phi^{-\frac{D-3}{D-2}}g^{ij}D_i A^\alpha \cdot D_j A_\alpha
+q\Phi^{-\frac{1}{2}}(\overline{\psi}_+ A\psi_+ +\overline{\psi}_-
A\psi_-)-M\Phi^{-\frac{1}{2(D-2)}}(\psi_+^+\psi_- +\psi_-^+\psi_+)
\]
\[
-\overline{\psi}_+ e^j{}_a\gamma^a_+
(i\partial_j  +i\omega_{+j}+qV_j)\psi_+
-
\overline{\psi}_- e^j{}_a\gamma^a_-(i\partial_j
 +i\omega_{-j} +qV_j)\psi_-
\]
\[
+\frac{i}{8}\Phi^{\frac{1}{2}}f_{ab}
(\overline{\psi}_+ \gamma_+^{ab}\psi_+ +\overline{\psi}_-
\gamma_-^{ab}\psi_-)
\]
which tells us that the $U(1)$ gauge field and the Yang--Mills fields
yield positive sources for the dilaton, while the kinetic
term of $A$ provides a negative contribution.
There is some ambiguity with regard to the contribution due to the fermions.
However, on--shell the fermion contribution adds up to a positive
term. Therefore, generically the fermions will contribute negative sources
to the dilaton.

This appearance of positive and negative source terms for the dilaton
apparently works for any value of $D$.
Of course, the negative terms are absent
in the ten--dimensional formulation
of superstring theory
since eleven--dimensional
supergravity does not contain elementary fermions
and Yang--Mills fields.

However, the case $D=4$ is peculiar due to the logarithmic IR divergence of the
electrostatic potential in 2+1 dimensions. In a linear approximation
the dilaton $\ln\Phi$ behaves like a massless scalar field
coupled to external sources,
whence finiteness of energy requires a vanishing dilaton charge:
\begin{equation}\label{cond}
\int d^2 \mbox{\bf x}\,\Big(\frac{1}{4}f_{ij}f^{ij}
+\frac{1}{8}
(V+Af)_{ij}^\alpha (V+Af)^{ij}_\alpha
-\frac{1}{4}
g^{ij}D_i A^\alpha \cdot D_j A_\alpha
\end{equation}
\[
-\frac{1}{2}q(\overline{\psi}_+ A\psi_+ +\overline{\psi}_-
A\psi_-)+\frac{1}{4}M(\psi_+^+\psi_- +\psi_-^+\psi_+)
+\frac{i}{16}f_{ab}
(\overline{\psi}_+ \gamma_+^{ab}\psi_+ +\overline{\psi}_-
\gamma_-^{ab}\psi_-)\Big)
=0
\]

This property is similar to charge neutrality of the Coulomb gas in
two dimensions and can be derived from
conformal invariance of the
partition function or independence of the arbitrary length scale entering
the definition of the electrostatic potential. It can also be inferred
from the fact that the retarded potential
$\frac{\Theta(t-r)}{2\pi\sqrt{t^2-r^2}}$
yields a finite dilaton for static configurations only if the
sources add up to zero.

The self--trapping mechanism encoded in (\ref{cond})
is superficially stable with respect to quantum effects,
since calculation of the 1--loop
effective potential in the linear approximation and
in presence of an ultraviolet
cutoff $\Lambda$ yields a
$\Lambda^3\cosh(\ln\Phi)$--type potential of the dilaton.
However, consideration of an effective potential in a nonrenormalizable
theory is rather speculative, and it is clear that like in the
four--dimensional theory with gravity a proper treatment of quantum effects
would imply genuine stringy considerations.

The physics behind (\ref{cond}) is more transparent than in the case
of the Coulomb gas, since particles and antiparticles contribute in
the same way to the dilaton: If a gauge boson excites a dilaton field
the divergence of the resulting energy density implies
pair production of adjoint scalars and fermions to restore an asymptotically
vanishing dilaton. Clearly, $M$ suppresses
the production of fermions relative to adjoint scalars.
Stated in another way: The adjoint scalar and light fermions screen
the dilaton charge of the gauge bosons.

The observed stability of the dilaton in this framework is not in
contradiction with unwinding of the fourth dimension, since in the
parametrization (\ref{KK}) the length $L$ of the range of $x^{D-1}$
determines the radius of the internal dimension rather than the
dimensionless dilaton $\Phi$. Rescaling $x^{D-1}$ to an angular variable
through the substitution $\Phi\to L^4\Phi$ shows immediately that
the reasoning above only implies stability of an arbitrary large internal
dimension against low energy excitations. Unwinding should
arise as a nonperturbative string effect rather than a property of a
low energy effective field theory with time--dependent couplings.

On the other hand, there is also another possibility, which I would prefer:
There is no unwinding of the fourth dimension, since it was never curled up.
Anything involved in the previous investigation was a lowest order
Kaluza--Klein parametrization, which only requires translational invariance
along
$x^3$, but there is no truncation of $x^3$ to a finite
interval.
This also implies absence of an energy gap: We were only
looking at the lowest edge of a continuum, approximated in the dual
supersymmetric formulation by saturated solitons.
The content of the three--dimensional string scenario is both remarkable
and simple: The low energy degrees of freedom and the supersymmetry
which we expect to inherit from string theory motivate considerations of
three--dimensional effective theories, but this strictly does not imply
compactification of seven dimensions.
It is pretty clear that the simple Kaluza--Klein parametrization
(\ref{KK}) employed here represents at best a crude approximation
to lowest order string theory, and we have to think more thorougly
about the geometry and meaning of the 3--manifolds involved in the
infrared limit of string theory.
It is a natural expectation that the role of the
3--manifolds will find an explanation
within the holographic theory of space--time
\cite{LS,tH}, as has been pointed
out already by Vafa, cf.\ \cite{Wi2}.

The self--trapping of the dilaton explained above clearly does not
depend on the particular embedding of 2+1 dimensions.
It thus turns out
that the prospects to get rid of a rolling dilaton through the
three--dimensional picture proposed
by Witten
are as excellent as the prospects to get rid of the cosmological
constant.
This scenario might open up an unexpected and interesting
window to string phenomenology
once we develop a better understanding of the
role and the dynamics of 3--manifolds in four--dimensional
quantum gravity,
and Witten's proposal clearly deserves further study.\\[2ex]
{\bf Acknowledgement:} I would like to thank E.\ Witten for discussions and
for hospitality at the Institute for Advanced Study. This work was
supported by a grant from the DFG.

\end{document}